\documentclass[a4paper]{article}

\usepackage[english]{babel}
\usepackage[utf8]{inputenc}
\usepackage{amsmath}
\usepackage{graphicx}
\usepackage[colorinlistoftodos]{todonotes}

\title{Analysis of Conditions favorable for Ball Lightning Creation}

\author{Herbert Boerner\\
55131 Mainz \\
Germany\\
he.boerner@googlemail.com}

\date{\today}

\begin{document}
\maketitle

\begin{abstract}
This report uses a few well documented cases of Ball Lightning (or BL for short) observations to demonstrate a correlation between BL and positive lightning, especially strong positive lightning. This allows to draw conclusions and predictions about future BL observations and the production of these objects in the laboratory. Contrary to many current BL theories, these objects can be created without direct contact to a lightning channel.  Very high electric fields appear to be essential for the creation, together with the proper temporal structure of the field. So far no experiments have been performed along the ideas presented in this report.
\end{abstract}

\section{Introduction}

Despite research efforts dating back nearly 200 years \cite{arago1838tonnere} , ball lightning remains one of the most enigmatic natural phenomena. Its rarity, and the unpredictability of its occurrence in space and time, together with the short duration and the limited range of visibility makes its controlled observation very unlikely. 
The lack of any evidence beyond the anecdotal reports – whose reliability he doubted- prompted the lightning researcher K. Berger to write a very skeptical report on BL \cite{berger1973kugelblitz} . Other researchers are less skeptical concerning the existence of BL as an independent atmospheric phenomenon \cite{rakov2003lighting} , but is nevertheless obvious that progress in understanding BL is very slow. It is hampered by the lack of detailed reports on observations and by the failure to create objects in the laboratory which display the full range of characteristics usually associated with BL \cite{rakov2003lighting}.
BL has been reported to originate in a number of different situations:

\begin{enumerate}
\item from lightning channels directly
\item  at the ground at or close to lightning impact points
\item at considerable distance from CG lightning impact points, completely unrelated to the lightning channel
\item near to or perhaps in aircraft during the flight
\end{enumerate}
Concerning the first case, a photo from the prairie meteorite network has been available since 1977 \cite{tompkins1975photographic}. Recently, under serendipitous circumstances Chinese researchers have been able to record BL on video, together with its spectrum \cite{cen2014observation}. The BL object was created at the impact point of a CG lightning, so for the many cases where BL of type 2 was observed also independent evidence exists. For the third kind of BL objects, so far no such evidence is available to support the numerous observer reports.
\paragraph{}
Observations of BL objects in aircraft are well-established, but they present perhaps the hardest problems for the explanation of BL creation. 
\paragraph{}
In this paper an analysis of several BL observations is presented which suggests that positive CG lightning, especially of very high strength,  displays a high probability to create BL objects, also of the type 3 mentioned above. This hypothesis is supported by a number of collections of BL reports, which show a correlation between BL and positive lightning from winter thunderstorms. Possible reasons for this correlation due to discharge processes are discussed and conclusions for controlled observations are drawn. The observation of multiple BL objects in these cases indicates that the conditions for the creation were unusually favorable. This gives valuable hints for the parameters to be used in laboratory experiments aiming at producing these objects in a controlled way. 

\subsection{Collections of BL reports
}
The first reliable collection of BL reports is due to Brand \cite{brand1923kugelblitz}. From more than 600 reports, he selected only cases with sufficient information and credibility. In his analysis of the remaining 215 reports he concludes that BL observations are well correlated with lightning activity, but that the relative frequency of BL was lower in summer and higher in winter: relative to the number of lightning flashes, winter thunderstorms (October-March) are producing more BL objects that summer thunderstorms (April-September). He also concludes that BL observations tend to occur towards the end of thunderstorms. \paragraph{}

Many of the BL collections published since Brand show the same correlation: winter thunderstorms produce more BL objects per lightning than summer thunderstorms \cite{piccoli2011statistical}, \cite{rayle1962nasa}, \cite{beaty2006percom}, \cite{smirnow1990}, \cite{grigorjew1989statanal}.
Reports from Europe and north America agree surprisingly well in the percentage of BL objects observed in the winter half-year, which is about 13 percent. Reports from eastern Europe are a bit lower in the percentage (5-7 percent) but still significantly above what could be expected from winter lightning, which is about 1-2 percent of the total lightning flashes per year based on the data of BLIDS and ALDIS \cite{blidsaldis}. It is well known that winter thunderstorms produce much less flashes than summer thunderstorms but they have a higher percentage of positive CG lightning. Positive lightning also occurs more at the end of thunderstorms. This information was of course unknown at the time when Brand wrote his book. With respect to BL research, the link between BL production and positive lightning was first mentioned in the book of Stenhoff \cite{stenhoff1999}.

\subsection{The Neuruppin case
}
An example of BL production in a winter thunderstorm is the Neuruppin case \cite{baecker2007multiple}, which is especially valuable since it is very well-documented. On January 15th 1994, the first and strongest lightning of a brief winter thunderstorm created a number of BL objects in the small town of Neuruppin. The lightning was a positive CG flash with an exceptional strength of 370 kA peak current as measured by the lightning detection system BLIDS (at 16:08 UTC). It produced a very strong illumination and an extremely loud thunder. It was followed by four less intense discharges between 16:09 UTC and 16:22 UTC, where the two last ones – one positive and one negative - were nearly coincident in time.  These subsequent flashes were further to east than the first one, the distances being compatible with a movement of the charge source of about 90 km/h, which is the speed by which the maritime air of polar origin was moving over Germany \cite{baecker2007multiple}. Other than these flashes, the thunderstorm produced only a cluster of negative flashes about 30 kilometers to the west near the town of Kyritz, thus showing a bipolar structure with respect to the CG lightning polarity.\paragraph{}

Immediately after the first extremely strong  lightning, people from the town Neuruppin phoned the local meteorological station to inquire about the nature of this event. The staff of the station collected the incoming reports and also asked for more reports of anything unusual connected with this lightning. To their surprise, many reports of BL objects were received. In total, 11 BL objects were reported, 2 large ones outside, but also several inside houses, and several ones crossing curtains or windows entering rooms from outside. Three BL objects were created inside the houses and could be observed by several people for a brief time. Taken together, the BL objects observed at Neuruppin display the full range of characteristics commonly attributed to BL: the sudden appearance, an irregular motion, passage though windows and curtains etc. Obviously, all these objects must have been created by the  same lightning discharge; therefore it also very likely that they all of the same nature.
The staff of the meteorological station collected all these reports and published them in an internal paper of the German meteorological service \cite{dwd1994sonder}. Less than a year later, two volunteers visited the witnesses again and collected more details about the observations. Ten years later, these reports were correlated with the data from the lightning detection network and the results were published \cite{baecker2007multiple}.\paragraph{}

The Neuruppin case is spectacular but not completely unique: a very similar case is mentioned in Brand as case number 55.  February 24th, 1884, in Amiens the first and only lightning of a winter thunderstorm -also one of exceptional strength- created 7 BL objects \cite{decharme1884orage}. It is likely that in this case the initiator was also a positive CG  lightning. Another similar case is mentioned in \cite{mathias1927foudre}. June 7th, 1925, in Pontgibaud one strong lightning created four BL objects.\paragraph{}

A very important fact is the distance between the lightning impact point and the BL observations: the detection system places the impact point more that 5 kilometers away from the center of Neuruppin.
This suggests that strong positive lightning of winter thunderstorms and maybe also from summer thunderstorms can create BL objects with surprisingly high probability at a considerable distance from the impact point. This is contrary to the common opinion that a direct interaction between lightning channel and BL object is  necessary. In the Neuruppin case, the creation of at least two BL objects inside houses was observed directly, they appeared suddenly and were directly observed in their final shape. Other observations (number 30 by Brand \cite{brand1923kugelblitz}, and also in 20.2.1 by Rakov and Uman \cite{rakov2003lighting}) also describe the sudden appearance of BL objects “out of thin air” without the direct interaction with a lightning channel. 
In addition to BL objects also corona discharge was  observed, which is useful to estimate the electric field generated by the flash. One observer looking towards the impact point of the lightning saw “huge blue bundles of flame extending towards the sky”, most likely negative streamers or leaders, indicating an electric field well in excess of 2-3 MV/m. This is roughly the value used in high voltage engineering for the breakdown field of negatively charged conductors \cite{kuechler1996hochspannungs}. This huge field was obviously sustained for an appreciable period of time, probably for several seconds. The distance from this observer to the calculated impact point was 4.7 km. The charge must have moved rather slowly at first, or it was not completely discharged by the first flash, since the corona discharges were present for a period of several seconds, which can be seen from the second observer who saw the corona before and after the thunder on a metal sieve used for sifting sand. This second observation of corona took place close to the region where also the BL objects were seen. Obviously also in this area the field was close to breakdown value.\paragraph{}

The BL objects were observed a bit  further away from the impact point (5-6 kilometers); here no streamers or corona discharges were reported but there were other indications of high electric fields, like the activation of a toy animal (which was controlled by metallic contacts on its head). It is therefore likely that also in the region of the BL objects the electric field was exceptionally high, even if the distance to the hit point of the lightning was several kilometers. This may have been due to the extreme strength of the lightning or to some unusual movement of the positive charge in the clouds.
Corona on a massive scale was also observed in Kyritz (where the cluster of negative flashes was centered), but BL objects were exclusively reported from Neuruppin. \paragraph{}

This creation of BL “at a distance” raises the question where the energy for the creation of such objects comes from. The only source of energy readily available and confirmed by observation is the electrical field created by the charge of the thunderstorm and the approaching lightning, but this is only a weak source of energy. The energy density of the electric field is proportional to the square of the field strength.  Therefore high fields up to the breakdown of air have a dis-proportionally high energy density, which may very well be essential for the creation of BL objects. For a field of 1 MV/m, the energy density is 4.4 J/m3, and for 3 MV/m, it is 39.8 J/m3. Stenhoff argues \cite{stenhoff1999} that most BL objects have only a small energy content of less than 3 kJ and Stepanow \cite{tepanow1997energybl} concludes that indoor BL has only up to 100 J, so at least the initiation of such objects can be envisaged via this source of energy. For BL objects outside, current flow through the object may be an additional source of energy \cite{stenhoff1999}.\paragraph{}

However, these high electric fields often lead to the local breakdown of air and the creation of streamers, which compete with the BL formation in terms of the available energy. It is likely that almost always streamers are produced and only in rare cases, where no streamers could form, BL objects are initiated. Case number 184 in Brands book \cite{brand1923kugelblitz} reports on such an event, where a strong lightning created large streamers everywhere in a village in France, except above a body of water where a BL object was created. The water provided a flat, conducting surface not suitable for the initiation of streamers. \paragraph{}
A very important fact is that positive CG lightning is fundamentally different from negative lightning with respect to the formation of streamers.
From laboratory experiments it is known that streamers from a negative point towards a positive plate (negative pre-breakdown streamers) need much higher fields (about 2 – 3 times) for their development than streamers starting from a positive point \cite{macgorman1998naturestorms}. The reason is that the mobile electrons have to move into regions with lower field. Positive lightning is thus more likely to produce a region around the impact point where the electric field is extremely high, providing a high energy density but not producing streamers which divert the available energy into breakdown processes. In this region, the energy density is 4 to 9 times higher than in the equivalent zone around negative CG lightning. It is most likely that this is the region where favorable conditions for the creation of BL objects exist.
The hypothesis that these very high electric fields are required for the initialization of BL objects also explains quite naturally why at the lightning research labs working with instrumented towers BL has not been observed so far \cite{berger1973kugelblitz}. Most of the lightning at these locations start from the towers upward, therefore the electric field at ground level never reaches the required high values. This is also the case for the lightning which originates from the clouds but is intercepted by the tower acting as a huge lightning rod.  Berger gives the maximum field strengths at ground level at his laboratory as 150 kV/m \cite{berger1973kugelblitz}, far below the breakdown fields mentioned above. At this field the energy density is only 0.1 J/m3.

\subsection{Possibilities for observation of BL objects
}

The hypothesis that positive CG lightning is more likely to produce BL objects than other types of lightning (e.g. negative CG lightning) allows predictions how to enhance the chances of observations. First, winter thunderstorms with strong positive lightning are obvious candidates. Using the data from the many lightning detection systems now in operation one could check if around the impact point of positive lightning something unusual had been observed. Other thunderstorms producing a high amount of positive lightning due to other reasons like smoke from forest fires \cite{rakov2003lighting} are also possible targets. \paragraph{}
Volcanic lightning is another likely choice. The plume of fine volcanic ash carried away by the wind has a positive charge \cite{thomas2007mtstaugustine}. There is one report, unfortunately only second hand and very brief, on the observation of numerous BL objects during the eruption of the Santa Maria volcano in Guatemala in October 1902 \cite{santamaria1905}. The observer was in San Cristobal Cucho \cite{cucho} about 29 kilometers away from the eruption vent and the volcanic ash fall there was very heavy, creating a layer of about 75-100 cm thickness. Since the eruption of the Santa Maria was one of the three strongest eruptions in the twentieth century with an eruption index of 6 on a scale of 8, this is obviously only an option for the more adventurous observers, but at least in this case, time and place of the BL creation are predictable. Also the conditions for BL creation must have been very favorable, since the report states that numerous BL objects have been observed, in this case coincident with strong corona discharge.\paragraph{}
Rocket triggered lightning \cite{rakov2003lighting} could perhaps also be used, but only if the wire trailing the rocket is not connected to ground and if the triggering is performed during a winter thunderstorm, where strong positive lightning may be induced by this technique. If the wire remains connected to ground the situation is more or less identical to the tower generated lightning, which would be unfavorable for the production of  BL objects.

\subsection{Characteristics of positive lightning
}

Besides the strong electric field and the reduced tendency to produce streamers, other characteristics of positive lightning may be important for the production of BL objects. Radio frequency pulses emitted by the lightning may be one of the essential ingredients, but unfortunately, nothing about the radiation emitted by the strong positive lightning flashes is known so far. \paragraph{}
Positive lightning is currently an important field of study in lightning research \cite{rakov2003positivebipolar}. Therefore it can be expected that a better understanding of their properties will also lead to more insight into the conditions which are required to produce BL objects by these lightning flashes.

\subsection{Parameters for creation of BL objects in the laboratory
}
Numerous attempts have been made to create BL objects in the laboratory. A good summary of the work until about 1980 is given in Barry's book \cite{barry1980bl}, or in Stenhoff's \cite{stenhoff1999}. Most of these experiments manage to create glowing regions of gas, but Uman and Rakov state that “none of these discharges exhibits the salient  characteristics of Ball Lightning, however.” \cite{rakov2003lighting}. In order to avoid discussions what these characteristics really are, one can simply compare the assumptions used in these experiments to the conclusions drawn above from the Neuruppin case. \paragraph{}
Many experiments assume a direct interaction of the lightning channel with ground. Recent examples are experiments that have been performed with the combustion of silicon \cite{abrahamson2000oxidation}. This is in clear contradiction to the observations at Neuruppin. 
Another large fraction of experiments work with discharges like sparks etc. Recently, experiments have been performed with discharges in water \cite{versteegh2008plasmoids}. These are also in contradiction with the observation of the BL objects appearing in closed rooms where no discharges have been observed.
Experiments working with combustion of flammable gases cannot explain the correlation between positive lightning and BL object production. 
  A number of experiments use high-power radio-frequency, but it is unclear if radiation of such strength and duration does indeed occur close to lightning strikes. This may, however, be different for very energetic positive lightning.
 From this comparison one can state that most if not all experiments performed so far require conditions which were not existing at Neuruppin. 
A more promising approach is to take the conditions observed at Neuruppin as a starting point and to reproduce them as closely as possible in the laboratory. The large number of BL objects created by one flash indicate that the conditions were unusually favorable for the formation. In the experiments, an electric field pulse should be applied to a test chamber. The following characteristics are most likely important:

\begin{itemize}
\item the correct polarity: positive electrode above
\item a complete avoidance of streamers from both electrodes
\item the correct temporal structure: negative charges from the electrode below should be able to travel a distance of up to one meter to be able to form a space-charge which allows the self-organization into a BL object.  This calls for a slow rise of the voltage which is of the order of tens of milliseconds rather than microseconds.
\item a peak field close to the breakdown of air: 2-3 MV/m
\item a shape of the negative electrode which helps to produce the correct shape of the space-charge 
\item the possibility to apply radio frequency pulses 
\end{itemize}
 
The requirement concerning the temporal structure is motivated by several facts. In the two cases in Neuruppin, where the creation BL objects were directly observed in houses, the luminous forms were seen to appear directly in their final shape “out of thin air”. The hypothesis is that a preexisting space charge of more or less correct shape is required for the formation of the BL object by the final, strong electrical pulse of the lightning. Since the mobility of small negative ions in air is about 10-4 m/sec / V/m \cite{gorman1998elenatstormsp33}, even in a strong field of 200 kV/m ions would need about 50 milliseconds to travel from a surface to a height of one meter.  If the field was higher than 250 kV/m, runaway electrons may have been present \cite{gurevich2005runaway} which would have moved much faster. The electrical field in Neuruppin may well have been at that level a considerable time before the actual lightning because one of the witnesses saw corona on a metal sieve before and after the lightning \cite{baecker2007multiple}. In \cite{schumann2012fieldchanges} it is stated that “Positive lightning flashes to ground are often preceded by significant in cloud discharge activity lasting, on average, more than 100 ms“. Such an intra-cloud activity may have been the source of an early charge generation at Neuruppin.\paragraph{}
Very puzzling is of course the fact that BL objects have never been observed in high voltage laboratories around the world. Most of these labs use generators like Marx Generators which create very high but only short voltage pulses, where the rise time is of the order of microseconds. These generators do not simulate the temporal structure of the electric field of an approaching leader \cite{gumley1}. On the other end of the temporal spectrum DC voltages are used. In between these two extremes is a range of rise times of the order of one to several hundred milliseconds, where no high voltage generator appears to be existing except for one developed for lightning protection research \cite{gumley2}.  Most likely this generator can produce the pulse shape required for BL experiments, but is used for testing lightning rods and therefore it neither works with the correct polarity nor is it being used in a situation where streamers have to be avoided completely. \paragraph{}
The requirements concerning the shape of the negative electrode are more difficult to establish. BL reports often state that the object was created above a flat, conducting surface. Examples are: above a 80 by 80 cm stone plate which perhaps was wet from the previous rain (Brand No. 30), above an iron stove (Brand No. 89), above a wet road (Uman 20.2.1), above a water surface (Brand No. 184). These cases may be due to the fact that flat surfaces inhibit the production of streamers. Other reports suggest that the shape of the conducting surface influences the size of the BL object created above and also its motion, for example in \cite{turner1996propbl}, where a BL object was observed above a large metal-covered table. In one case, BL objects were repeatedly observed close to an iron stove pipe \cite{rockbetu}. \paragraph{}
The influence of radio frequency pulses on the creation of BL objects is unclear, but it should be noted that the conditions mentioned above regarding the shape of the negative electrode could also indicate that some form of crude EM resonator is essential.

\subsection{Summary}
The Neuruppin case conclusively demonstrates that BL is frequently created at a distance from the lightning channel. No direct contact between the lightning and BL is required. The energy of these objects must therefore come from the electric field of the lightning, at least in the initial stage.\paragraph{}
Positive CG lightning produces less breakdown processes in form of streamers than negative lightning, leading to higher field strengths above ground and therefore to much higher energy densities of the field. Compared to negative CG lightning, there is a larger area around the impact point of positive lightning where BL objects may be created. The Neuruppin case demonstrates that under these circumstances a multitude of these objects can be created by one lightning, indicating very favorable conditions for the formation.\paragraph{}
This hypothesis provides a possible explanation for the correlation between positive CG lightning and BL object creation. It also gives an explanation for the negative results for tower generated lightning and rocket triggered lightning with respect to BL observations.\paragraph{}
Better chances for BL observation exist around impact points of strong positive lightning in winter thunderstorms or other thunderstorms with a high percentage of positive lightning and also for lightning due to strong volcanic eruptions.
The ideas developed above cannot explain BL creation inside flying aircraft. Modern aircraft are good Faraday cages, so their interior is very well shielded from electric fields.  Only EM radiation can penetrate into the interior via windows or antenna feedthroughs. It is, however, not clear if the BL objects observed inside aircraft were not always formed outside and entered the aircraft via e.g. the front windows \cite{stenhoff1999}. \paragraph{}
The hypothesis that positive lightning produces BL objects with high probability allows to define at least some of the parameters required to produce these elusive objects in the laboratory. It is advisable to use these conditions as a starting point for further experiments.\paragraph{}
For a better definition of the parameters field studies should be performed to investigate the conditions created by strong positive lightning, especially the radio frequency pulses emitted and the temporal structure of the electric field.
In general, more work on the exact circumstances of BL object creation is required. This should encompass an analysis of all reports where the initiation of these objects has been observed directly.  It is likely that other possibilities may be found to create these enigmatic objects in the laboratory enabling at long last a thorough scientific study of Ball Lightning. 

\subsection{Acknowledgment}

The author thanks Dr. A. Keul for his help in retrieving references \cite{beaty2006percom}, \cite{decharme1884orage} and \cite{mathias1927foudre}

\end{document}